\documentclass[aps,pra,10pt]{revtex4-2}
\usepackage{graphicx}
\usepackage[utf8]{inputenc}
\usepackage{amsmath, amsthm, amsfonts}
\usepackage{dcolumn}
\usepackage{bm}

\usepackage{lineno}
\usepackage{orcidlink}

\newcommand{\bracket}[1]{\left\langle #1\right\rangle}

\newcommand{\be}{\begin{equation}}
\newcommand{\ee}{\end{equation}}
\newcommand{\bd}{\begin{displaymath}}
\newcommand{\ed}{\end{displaymath}}

\newcommand{\beeq}[1] {\begin{equation}\begin{split}#1\end{split}\end{equation}}

\begin{document}
\title{Understanding and improving axial detection in optical tweezers based on the interference of forward- and backward- scattered light}

\author{Isaac P\'erez Castillo$^1$\orcidlink{0000-0001-7622-9440}, Simon Leturcq$^2$, Sylvain Domitin$^2$, Ashley L. Nord$^2$\orcidlink{0000-0001-6199-2769}, Francesco Pedaci$^2$\orcidlink{0000-0001-6199-2769} and Alejandro V. Arzola$^{3,*}$\orcidlink{0000-0002-4860-6330}}

\affiliation{$^1$Departamento de Física, Universidad Autónoma Metropolitana-Iztapalapa, San Rafael Atlixco 186, Ciudad de México 09340, México}

\affiliation{$^2$CBS Un.Montpellier, CNRS, INSERM, Montpellier 34090, France}

\affiliation{$^3$Departamento de Física Cuántica y Fotónica, Instituto de Física, Universidad Nacional Autónoma de México, C.P. 04510, Cd. de México, México}

\affiliation{$^*$Author to whom any correspondence should be addressed.}

\email{alejandro@fisica.unam.mx}

\begin{abstract} 
Fast and accurate 3D position detection in optical tweezers (OT)  is essential for quantitatively monitoring subtle variations in the mechanical properties of microscopic systems ranging from biomolecules to cells and colloids. Because standard OT configurations do not provide direct access to the axial position, axial detection typically relies on temporal fluctuations in forward-scattered optical power to infer the position of the particle. This approach generally assumes a linear-response regime in which the signal arises from the interference between the forward scattered and the nonscattered optical fields; however, under certain conditions, the backward-scattered contribution becomes non-negligible, leading to deviations from the linear response.
Here, we present a simple yet comprehensive model for axial detection in standard OT while explicitly accounting for the backward-scattered field. Together with experimental validation, this framework neatly explains the standing-wave response observed when the backward-scattered field interferes with the nonscattered and the forward-scattered components, enabling accurate estimation of trap stiffness and particle diffusion under more general conditions. This work deepens our understanding of the phenomenology observed in real optical-tweezers measurements and extends their capabilities to conditions where standard approaches fail.     
\end{abstract}
\keywords{optical tweezers; calibration; position detection}

\maketitle

\section{Introduction}

Optical tweezers are widely used in physics and biophysics to manipulate microscopic objects and measure minute forces in systems ranging from biomolecules to cells and colloids \cite{ashkin1986observation, neuman2004optical, pesce2020optical, Gieseler:21,volpe2023roadmap}. Numerous experimental strategies have been developed to precisely track particle motion, reconstruct forces, and infer the rheological properties of the surrounding medium \cite{neuman2004optical, pesce2020optical, Gieseler:21,pralle1998local, tassieri2015microrheology, tassieri2019microrheology}. In particular, detecting the light scattered by the trapped particle using a photodetector enables an indirect determination of the particle's position and the forces acting on it, with the achievable accuracy set by the linearity and sensitivity of the detection scheme.

A standard optical-tweezers setup is typically built on a conventional inverted microscope, in which the objective lens generates the optical trap while the condenser collects the transmitted light to monitor the particle's position at high sampling rates \cite{gittes1998interference,florin1998photonic}. For lateral ($x-y$) position detection, a position-sensitive detector or a quadrant photodiode  measures the lateral displacement of the forward-scattered component of the laser beam. In contrast, the axial position, $z$, is inferred from temporal variations of the total signal (optical power)  collected by the detector. In this configuration, since the detection of $z$ relies on low spatial frequencies of the collected field,  one must trade off the resolution between $x-y$ and $z$ \cite{samadi2011role}. An iris diaphragm is commonly used to make this compromise. A more sophisticated configuration uses  an independent photodiode to detect position $z$, allowing a high-numerical-aperture consender for $x-y$ detection and a separate small apertured channel  for $z$ \cite{samadi2011role, dreyer2004improved}. The latter is used in this work to study the detection in $z$. 

It is commonly assumed (and often desired) that the scattered field produces a  photodiode signal  linearly proportional to the displacement of the trapped particle, $s(t)-s_0=\beta z$, where $\beta$ is the position-to-voltage conversion factor or sensitivity, and $s_0$ is the mean baseline voltage. This linear relationship  arises solely from interference between the outgoing laser beam and the forward-scattered field \cite{gittes1998interference,florin1998photonic, pralle1999three, rohrbach2002three}. However, under certain experimental conditions, the backward-scattered field  reflected by the bottom cover glass becomes non-negligible. The backward-scattered light reflected by the bottom cover glass contributes to the total interference field  collected by the condenser lens, producing a standing-wave effect along the axial direction ($z$), with a period equal to half the wavelength of the trapping laser in the medium. Because this contribution is very small, this field has been shown to not contribute to optical forces, affecting only the detection sensitivity \cite{neuman2005measurement}. This contrasts with the standing-wave phenomenon that forms when the particle is axially moved towards an interface with refractive index mismatch, where the interference of the outgoing laser beam with its reflection contributes substantially to the deformation of the energy potential even when the reflectivity is low \cite{jonavs2001single, jakl2003behaviour, soni2013periodic}. 

Backward scattered light can affect the accuracy of position detection in photonic force microscopy, particularly for surface imaging beyond the diffraction limit \cite{desgarceaux2020high, friedrich2015surface}. Due to its non-linear dependence on position, this contribution is often neglected in conventional optical tweezers experiments \cite{pralle1998local, florin1998photonic, hansen2005novel}. However, it has been used to determine the effective focal shift of a focused beam in sample chambers exhibiting spherical aberrations \cite{neuman2005measurement, schaffer2007surface, schaffer_surface_2007}. Despite its ubiquity, this phenomenon remains largely unexplored and lacks a comprehensive theoretical description. 

In this work, we present a comprehensive model that accurately describes the axial position, forces, and statistical properties of optical tweezers directly from the photodiode signal. The model explicitly accounts for the standing-wave behaviour arising from the backward-scattered field reflected by the bottom cover glass. We derive key statistical quantities, including the autocorrelation function and power spectrum density, and we experimentally demonstrate that the model accurately captures the observed phenomenology. 

\section{Interference detection with backward- and forward-scattered light}
\begin{figure}[ht!]
    \centering
\includegraphics[width=0.6\linewidth]{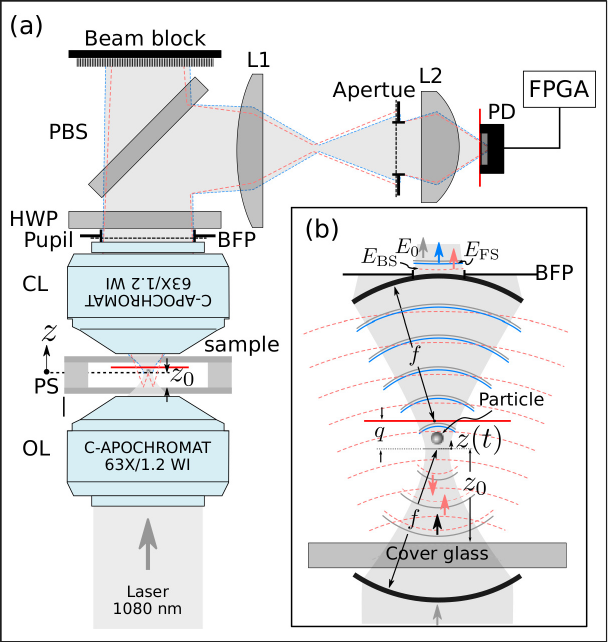}
        \caption{ (a) Experimental configuration and axial ($z$) position detection in an OT. A linearly polarized Gaussian beam is tightly focused from below using a water-immersion objective lens (OL), trapping a particle near the beam waist inside the sample chamber. The outgoing light is collected from above by the lens condenser (LC). The low-spatial-frequency components of the collected field pass through the pupil (Pupil), located at the back focal plane (BFP), and the transmitted light is then relayed by means of a half-wave plate (HWP), a polarizing beam-splitter (PBS) and the lens L1 toward the aperture stop (Aperture), after which it is focused by lens L2 and detected with a photodiode (PD). Red solid lines in the sample and in the detector indicate conjugated planes of detection. The resulting signal is digitized using a data-acquisition card (FPGA). The sample cell (Sample) is mounted on a piezoelectric stage (PS) which controls the particle's axial position relative to the cover glass according to $z_0=z_{\rm stage}-\delta_z$, where $\delta_z$ is a constant offset.  (b) Schematic representation of light propagation. The forward-scattered field (blue wavefronts) and the backward-scattered field (red wavefronts, reflected by the bottom cover glass) interfere with the outgoing laser beam at the conjugated plane of detection indicated by the red solid line, which is located at the focal point of the LC, at distance $q$ from the focal point of the OL. Here, $z(t)$ denotes the instantaneous particle displacement relative to its equilibrium position near the beam waist, and $z_0$ is the distance from the cover glass to the equilibrium position. The technical details of the setup are presented in Methods and materials in the Supplementary information.}
    \label{fig:scheme}
\end{figure}

A silica particle is trapped by a tightly focused laser beam (wavelength in vacuum $\lambda_0=1080\,n\mathrm{m}$) in an inverted microscope configuration, as schematically illustrated in Fig.~\ref{fig:scheme}.  The particle is suspended in water at room temperature ($T=22^oC$). The objective lens (OL) and the condenser (CL) are water-immersion objectives with a numerical aperture $\mathrm{NA}=1.2$. While the objective focusses the trapping beam, the condenser collects the low spatial-frequency components of the transmitted light. Water-immersion objectives allow focusing and collection without introducing spherical aberrations. The optical power of the collected field is then relayed and integrated by a photodiode. In other implementations, the beam block shown schematically in Fig.~\ref{fig:scheme} can be removed and replaced with a position-sensitive detector for detection of $x-y$.  

\begin{figure*}[ht!]
    \centering
\includegraphics[width=1\linewidth]{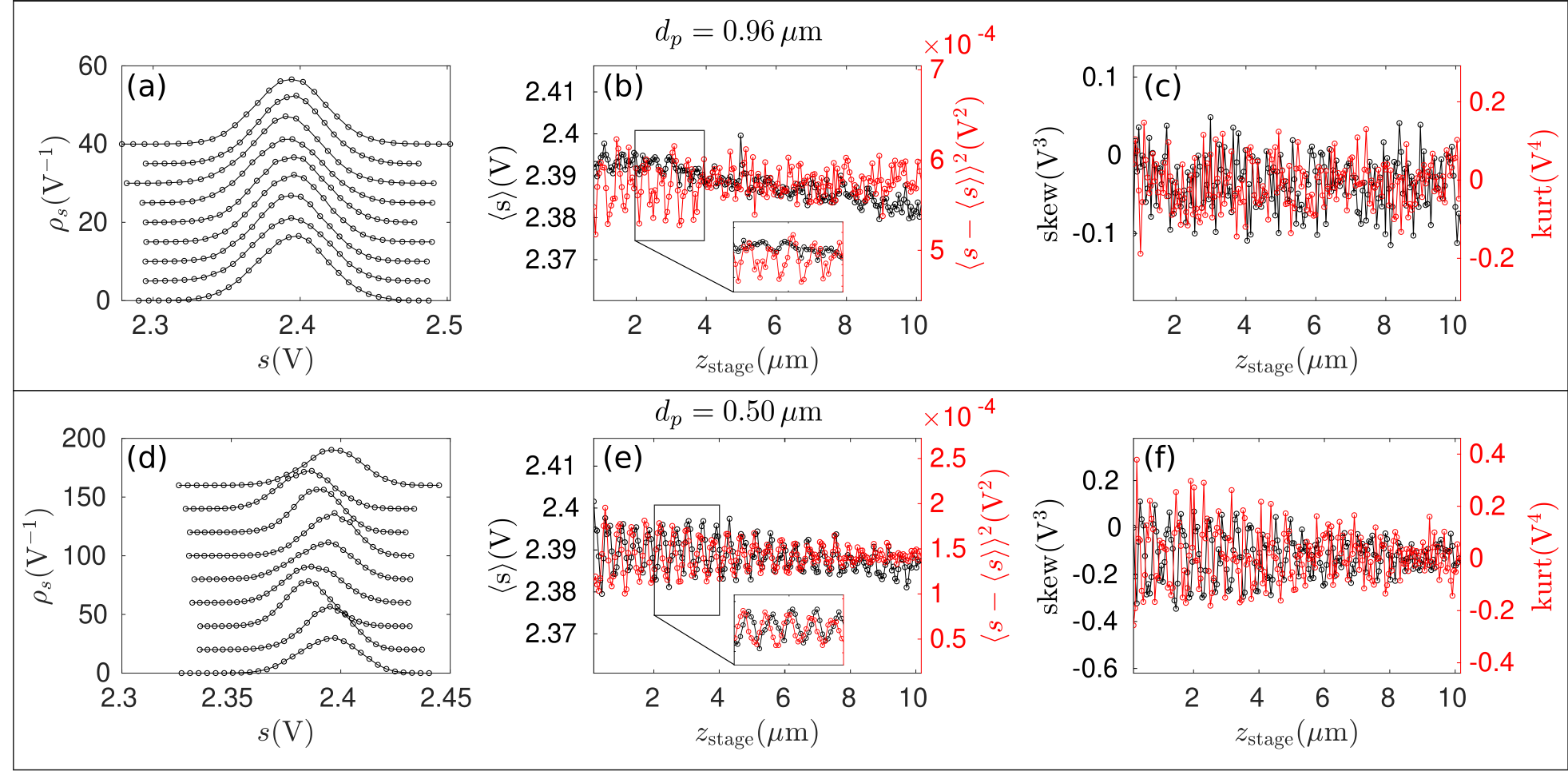}
    \caption{Statistical behavior of the photodiode signal as a function of the distance between the trapped particle and the bottom cover glass for particles with diameters $d_p=0.96\,\mu\mathrm{m}$ (a-c) and $0.50\,\mu\mathrm{m}$ (d-f). The particle position $z_{\rm stage}$  approximately corresponds to the distance between the particle equilibrium position and the cover glass $z_0$. (a) and (d) show the probability density function ($\rho_s$) of the signals ($s(t)$) at several evenly spaced positions $z_{\rm stage}$, ranging from $z_{\rm stage}=2.2\,\mu\mathrm{m}$ (bottom) to  $z_{\rm stage}=3\,\mu\mathrm{m}$ (top). (b) and (e) display the mean and variance of $s(t)$ as a function of $z_{\rm stage}$. Insets in these figures show the zoomed out plots in the range $z_0=(2-4)\mu\mathrm{m}$. (c) and (f) show the corresponding skewness and kurtosis. Both particles were trapped using the same laser power $P=25 \pm 1 \,m\mathrm{W}$, estimated inside the sample.}
    \label{fig:signals}
\end{figure*}

The data in Fig.~\ref{fig:signals} illustrate the main statistical properties of the detected signal ($s(t)$) —its probability distribution, mean, variance, skewness and kurtosis— measured for large and small silica particles ($0.96\,\mu \mathrm{m}$ in (a-c) and $0.50\,\mu\mathrm{m}$  in (d-f)) at different distances $z_{\rm stage}$ from the cover glass, with $z_{\rm stage}$ corrected to  approximate the particle-cover glass separation $z_{\rm stage}\approx z_0$. The first-column panels (Figs.~\ref{fig:signals}(a,d)) show how the signal distribution changes with $z_{\rm stage}$ over approximately one wavelength in the medium ($\lambda=814.79\,n\mathrm{m}$ using the refractive index of water $n=1.3255$). Whereas the distributions for the larger particle remain largely invariant with $z_{\rm stage}$, those for the smaller particle exhibit pronounced variations, alternating between narrower and broader profiles depending on position. Both the width and the mean of the distributions oscillate with $z_{\rm stage}$. This phenomenon can be attributed to the interference among the forward- and backward-scattered fields and the unscattered component, as  described below.

These oscillations are seen more clearly in Figs.~\ref{fig:signals}(b,e), which plot the mean (black scale) and variance (red scale) of the signal as functions of $z_{\rm stage}$. The oscillations are more pronounced in the variance —even for the larger particle— and one also observes that the variance and mean curves are phase-shifted by $\pi/2$ (see the zoomed inset in Fig.~\ref{fig:signals}(e)). In general, $s(t)$ is not normally distributed: its skewness and kurtosis deviate from zero, as shown in Figs.~\ref{fig:signals}(c,f). For the larger particle, both quantities remain close to zero, indicating an approximately normal distributed signal,  whereas for the smaller particle, they display a biassed oscillatory behaviour. We attribute this to the non-linear dependence of $s(t)$ on $z$ and the oscillatory dependence on $z_{\rm stage}$. This demonstrates that the commonly assumed linear relationship between signal and position fails, so a more accurate theoretical description of the phenomenon is required. 

\section{Position to voltage signal conversion}
 We assume that the detected field arises from interference among the unscattered incident field ($\tilde{E}_0$), the forward-scattered field ($\tilde{E}_{\rm FS}$) and the backward-scattered field ($\tilde{E}_{\rm BS}$) reflected from the bottom interface, as sketched in Fig.~\ref{fig:scheme}(b). Under a scalar-field approximation,
 the intensity amplitude at the detection plane is
\beeq{\label{eq:interference}
|\tilde{E}_{\rm BFP}|^2&\approx |\tilde{E}_{0}|^2+
2|\tilde{E}_{{0}}||\tilde{E}_{\rm FS}|\cos(\theta_{0}-\theta_{\rm FS})\\
&+2|\tilde{E}_{0}||\tilde{E}_{\rm BS}|\cos(\theta_{0}-\theta_{\rm BS}),
}
where we neglect $|\tilde{E}_{\rm FS}|^2$, $|\tilde{E}_{\rm BS}|^2$ and $2|\tilde{E}_{\rm FS}||\tilde{E}_{\rm BS}|\cos(\theta_{\rm FS}-\theta_{\rm BS})$. Because the particle's displacement within the optical trap is small compared with the characteristic axial scale of intensity variation, $|\tilde{E}_{\rm BFP}|^2$ varies in $z$ primarily through the interference terms (the second and third terms). The amplitudes $\tilde{E}_0$ and $\tilde{E}_{\rm FS}$ are independent of the cover-glass distance, whereas $\tilde{E}_{\rm BS}$ depends on it. From simple geometric considerations, the amplitude of the back scattered field evaluated at the detection plane is proportional to the inverse of the distance, $|\tilde{E}_{\rm BS}| \propto1/(2z_0+q)$, where $z_0$ is the distance between the particle's equilibrium position and the cover glass and  $q$ is the distance of the detection plane (condenser's focal plane) to the equilibrium position. This $1/r$-type scaling is also consistent with the field radiation or with the amplitude of a spherical wave. 

The phase of the incident field, $\theta_{\mathrm{0}}$, is constant, whereas the phase of the forward-scattered field, $\theta_{\rm FS}$, varies with the position of the particle in the focus due to the Gouy phase \cite{gittes1998interference, florin1998photonic}. Similarly, the phase of the backward-scattered field, $\theta_{\rm BS}$, depends both on the position of the particle with respect to the focal point and on its distance from the cover glass. If we assume a small  particle  and  a focused field well described by Gaussian optics, the phase difference between incident and forward-scattered fields is governed by the Gouy phase shift at the position of the particle $\theta_{0}-\theta_{\rm FS}\approx\mathrm{atan}(z/z_R)-\pi/2$, where the particle's equilibrium position is at the beam waist  and $z_R$ denotes the Rayleigh range (one-half of the depth of focus) of the focused beam \cite{saleh2019fundamentals}. For an ideal Gaussian beam $z_R=\pi w_0^2/\lambda$, where $w_0$ is the waist radius \cite{saleh2019fundamentals}. Taking $w_0=\lambda_0/\pi \mathrm{NA}$ gives $z_R=316\,n\mathrm{m}$.  Since the particle's displacement (typically tens of nanometres) is usually small  compared to $z_R$, this phase can be linearised around the equilibrium position, yielding $\theta_{0}-\theta_{\rm FS}\approx z/z_R-\pi/2$, and therefore $\cos(\theta_{0}-\theta_{\rm FS})\approx  z/z_R$. This recovers the familiar linear relation between voltage and position used in the literature, $s(t)-s_0=\beta z$, where $\beta$ is proportional to the interference power in the detection plane,  $\beta\propto2\int\int _{\sigma}(|\tilde{E}_0||\tilde{E}_{\rm FS}|/z_R )\mathrm{d}x\mathrm{d}y$ \cite{pesce2020optical, neuman_optical_2004}.    

Similarly, the phase delay between the incident field and the backward-scattered field is $\theta_0-\theta_{\rm BS}=z/z_R-\pi/2+4\pi z/\lambda+4\pi z_0/\lambda+\phi_0$, where $\phi_0$ is an effective, constant, phase shift that accounts for additional phase delays arising from fixed experimental conditions (e.g., finite particle size or an equilibrium position offset from the beam waist due to radiation pressure). The time-dependent power measured at the photodiode can then be written as
\beeq{\label{eq:powersignal}
P(t)\propto s(t)\approx s_0+\beta z+\frac{\gamma_0}{2z_0+q} \sin\left(w z(t)+\frac{4\pi}{\lambda} z_0+\phi_0\right),
}
where $\omega\sim 1/z_R+4\pi/\lambda$, and $\gamma_0/(2z_0 +q)\propto 2\int\int _{\sigma}|\tilde{E}_{0}||\tilde{E}_{\rm BS}|\mathrm{d}x\mathrm{d}y$. In general, the particle is not very small, and the beam does not follow the ideal Gaussian optics. In this case $z_R$ is not the paraxial Rayleigh range; however, one can define a constant effective range $z_{R,\rm{eff}}=\lambda/(\lambda \omega-4\pi)$ that tends to the Rayleigh range when the particle diameter is much smaller than the wavelength. The phase $\phi_0$ also includes an unknown constant delay due to the size of the particles. The stochastic dynamics of the trapped particle along the axial direction, $z(t)$, is governed by the restoring force of the optical trap with stiffness $\kappa$, the viscous drag characterised by the coefficient $\eta$, and thermal fluctuations (see Supplementary information). In the stationary regime, $z(t)$ is normally distributed with variance $\sigma^2=k_BT/\kappa$, where $k_BT$ is the thermal energy of the bath. 

In contrast to the common approximation that neglects the third term in Eq.~\eqref{eq:powersignal}, the distribution of the detected signal $s(t)$ strongly depends on $z_0$ and, in general, is not normal.  The general expressions for the mean and variance of the signal, to  first order in $\gamma$, are (see supplementary information for details of the derivations)
\beeq{\label{eq:meanSnonlin}
    \langle s (t;z_0)\rangle &=s_0+\gamma(z_0) e^{-\frac{1}{2}\omega^2\sigma^2}\sin\left(4\pi z_0/\lambda+\phi_0\right),}
\beeq{\label{eq:varsigma1}
\langle s (t;z_0)-\langle s (t;z_0)\rangle^2&=\beta^2\sigma^2+2\beta\gamma(z_0) \sigma^2\omega e^{-\frac{1}{2}\omega^2\sigma^2}\cos\left(4\pi z_0/\lambda+\phi_0\right)+\mathcal{O}(\gamma^2),
}
where $\gamma(z_0)=\gamma_0/(2z_0+q)$. Qualitatively, these expressions reveal an oscillatory dependence of both the mean and variance of $s(t)$ on the distance $z_0$, with a period $\lambda/2$, in agreement with the experimental results shown in Fig.~\ref{fig:signals}. The multiplicative factor $\gamma(z_0)$ explains the increase in the amplitude of the oscillation as the particle approaches the bottom cover glass. These expressions are also consistent with the phase shift $\pi/2$ between the mean and variance observed in the experimental data. The exponential factor in both equations depends on the positional variance, which is proportional to the inverse of the stiffness, implying that stiffer traps exhibit the strongest oscillations, whereas oscillations in weak (low-power) optical traps are attenuated. All these aspects will be characterised in more detail by analysing the power spectral density (PSD) of the signal. Finally, note that the data in Fig.~\ref{fig:signals} were acquired under similar conditions, using the same optical power and detection settings.

\section{Estimation of the diffusion and the stiffness}
The power spectral density (PSD) of the position of the particle is a standard tool for estimating both the stiffness of the trap and the diffusive (viscous) properties of the surrounding fluid \cite{pesce2020optical}. Alternative statistical estimators-such as the autocorrelation function or  maximum-likelihood methods-can also be used \cite{Gieseler:21, Perez-Garcia:23}. In this work, we focus on the PSD-based analysis, though the concepts developed here readily extend to other approaches. 

To derive the PSD of the detected signal, we first obtain the autocorrelation function. 
To first order in $\gamma$, and considering the long-term regime, the autocorrelation function is given by (see Supplementary information for details) 
\beeq{\label{eq:acf}
\mathrm{ACF}_s(t)&=\left[\beta^2+2\beta\gamma\omega e^{-\frac{1}{2}\omega^2 k_B T/\kappa}\cos\left(\frac{4\pi}{\lambda} z_0\right)\right]\frac{k_BT}{\kappa} e^{-t/\tau}+\mathcal{O}(\gamma^2),
}
where $\tau=\gamma/\kappa$ is the characteristic time of the optical trap. Fourier transforming this expression, neglecting higher-order terms, yields the power spectral density, 
\beeq{\label{eq:psdmodel}
 \mathrm{PSD}_s(f)&\approx\left[\beta^2+2\beta\gamma\omega e^{-\frac{1}{2}\omega^2 k_B T/\kappa}\cos\left(\frac{4\pi}{\lambda} z_0\right)\right] \frac{1}{2\pi^2}\frac{D}{f_c^2+f^2},
}
where $f_c=1/2\pi\tau$ is the cutoff frequency. Aside from the oscillatory modulation of the  amplitudes, for fixed $z_0$,  the ACF retains the characteristic  exponential decay and the PSD retains the Lorentzian form expected from linear response. The PSD amplitude, $A=(\beta^2+2\beta\gamma\omega e^{-\frac{1}{2}\omega^2k_BT/\kappa}\cos(4\pi z_0/\lambda)) D/2\pi^2$, oscillates with $z_0$, whereas the cutoff frequency $f_c$ is unaffected by these oscillations. Due to hydrodynamic interactions between the particle and the nearby cover glass, the diffusion can be approximated by Fax\'en's correction (to  third order)\cite{leach2009comparison}, 
\beeq{\label{eq:faxen}
D(z_0;d_p)=D_0\left(1-\frac{9}{16}\frac{d_p}{z_0}+\frac{1}{16}\frac{d_p^3}{z_0^3}\right),
}
where $D_0=k_BT/\eta$ is the bulk diffusion coefficient, $\eta=3\pi\nu d_p$ is the Stokes drag, and $\nu$ is the fluid viscosity ($\nu=0.95\times10^{-3}\,Pa s$). Here, $z_0$ denotes the distance between the centre of the particle and the cover glass. This size- and position-dependent diffusion affects not only the PSD amplitude but also the cutoff frequency, since  $f_c=\kappa D(z_0;d_p)/(2\pi k_BT)$. As shown below, this dependence enables a complete calibration of the optical-tweezers system.  

\begin{figure}[ht!]
\centering
\includegraphics[width=0.7\linewidth]{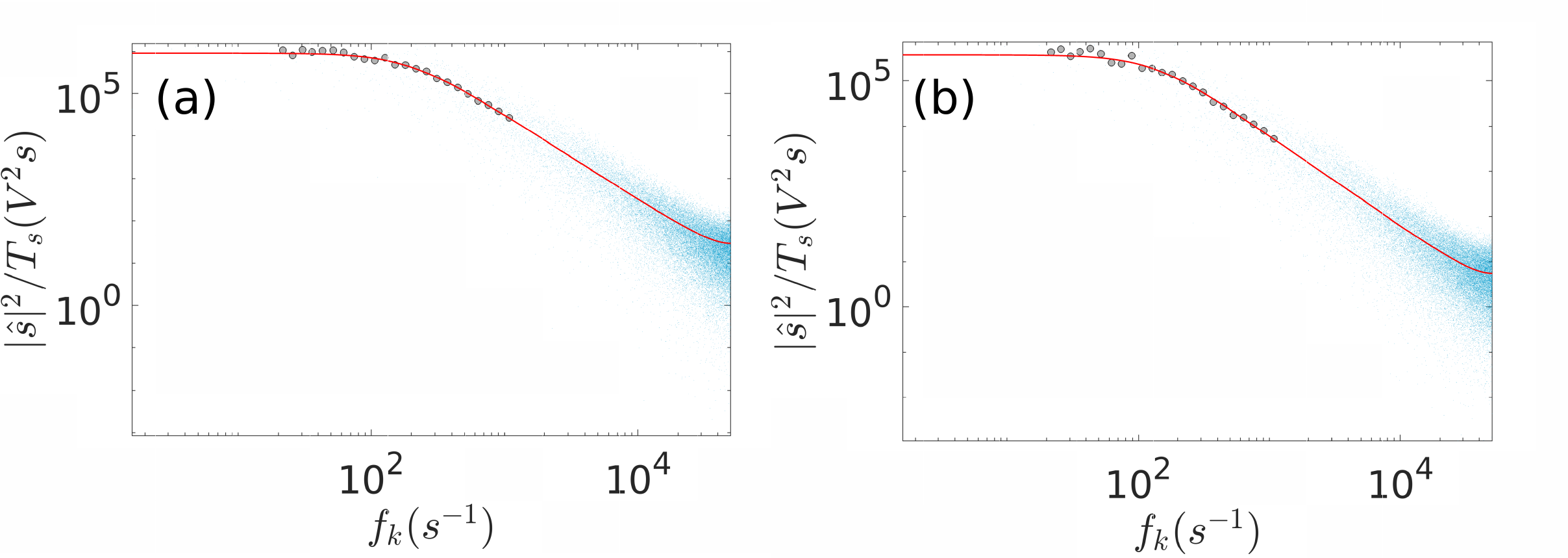}
    \caption{Examples of the power spectral density of the photodiode signal for particles of diameter $0.96\,\mu\mathrm{m}$ (a) and $0.50\,\mu\mathrm{m}$ (b). In each plot, blue dots correspond to $|\hat{s}|^2/T_s$, circles correspond to the expected values of the PSD at low frequencies, in the range $(20-1000) \mathrm{Hz}$, and red line corresponds to the Lorentzian fitting. The particles were located at position $z_{0}\sim 5\,\mu\mathrm{m}$ far from the cover glass, marked with red arrows in Fig.~\ref{fig:Afcsignal}. Experimental parameters are the same of those in Fig.~\ref{fig:signals}.}
    \label{fig:PDFs}
\end{figure}

\begin{figure}[ht!]
\centering
\includegraphics[width=0.7\linewidth]{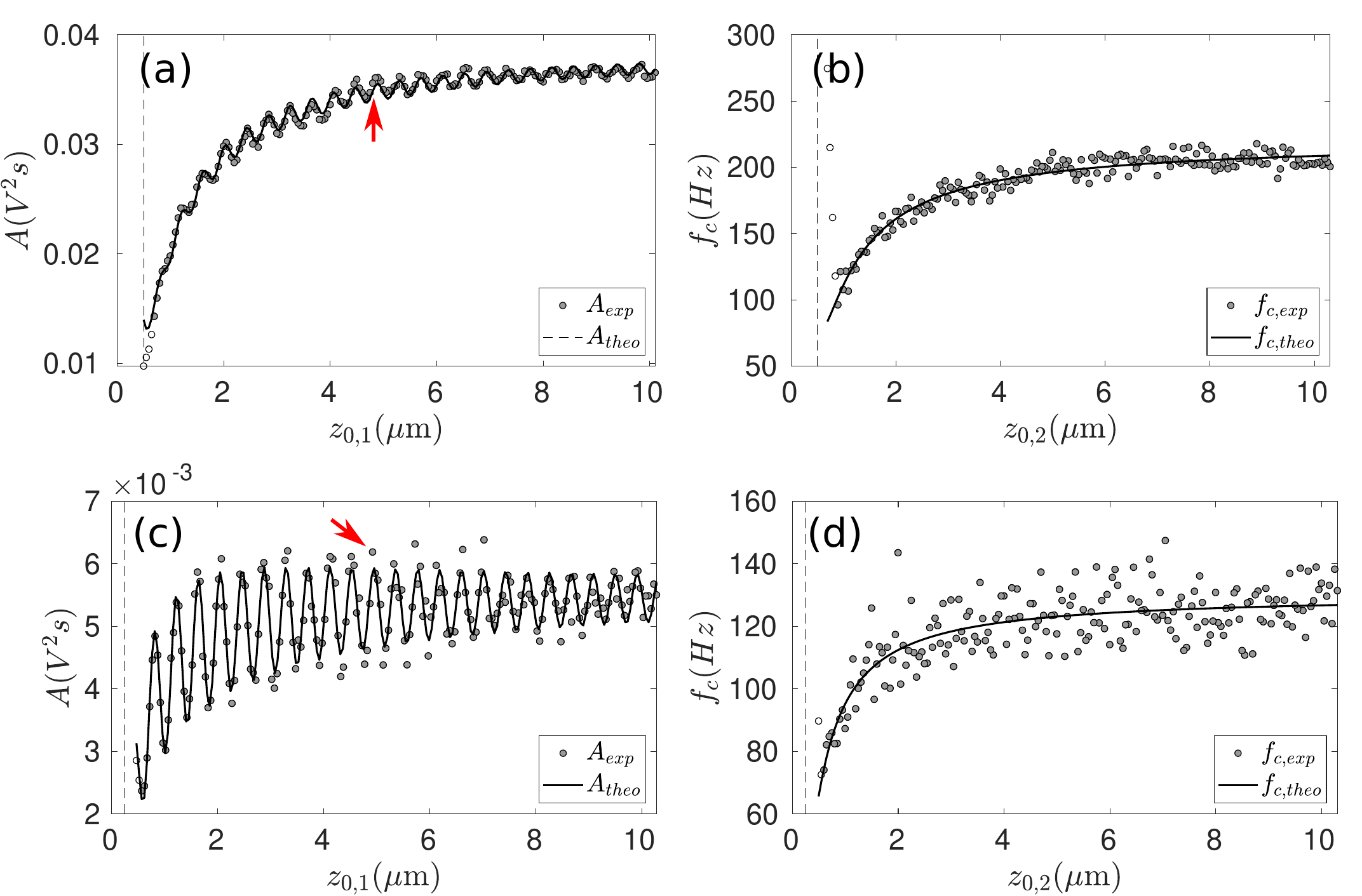}
    \caption{Amplitude ($A$) and cutoff frequency ($f_c$) of the PSD of the photodiode signal for trapped particles with diameters  $d_p=0.96\,\mu\mathrm{m}$ (a,b) and  $0.50\,\mu\mathrm{m}$ (c,d). The dots represent  experimental estimates obtained from nonlinear least-squares fits of the measured PSD to the Lorentzian model given by Eq.~\eqref{eq:acfpsd}, as shown in Fig.~\ref{fig:PDFs}. Arrows indicate data points corresponding to the fittings in Fig.~\ref{fig:PDFs}. The solid dark lines correspond to fits using the theoretical model described by Eq.~\eqref{eq:psd_amplitudefc}. The fitting results are summarized in Table~\ref{tab:tablefit}.}
    \label{fig:Afcsignal}
\end{figure}

In practice, to estimate the expected PSD of the experimentally acquired signal $\{s_n\}$, sampled at discrete times $t_n=n\Delta t$, with $n=1,2,...$, we first compute the periodogram as $|\mathrm{FFT}\{s_n\}(f_k)|^2/T_s$, where $T_s$ is the total acquisition time and $f_k=k/T_s$,  with $k=1,2,...$, is the sampled frequency. The resulting scattered spectrum is then averaged in logarithmically spaced frequency bins to improve statistical sampling \cite{berg-sorensen_power_2004}. The expected PSD is shown as gray circles in Fig.~\ref{fig:PDFs}. These data are then fit to a Lorentzian $A/(f_c^2+f^2)$ with aliasing corrections as in Ref. \cite{berg-sorensen_power_2004}, treating $A$ and $f_c$ as free parameters. The resulting fits are shown as red lines in Fig.~\ref{fig:PDFs}. The fitted values of $A$ and $f_c$ as functions of the distance from the cover glass are shown as gray circles in Fig.~\ref{fig:Afcsignal}. The solid lines in Figs.~\ref{fig:Afcsignal}(a, b) correspond to fits of the theoretical amplitude, $A_{theo}=(\beta^2+2\beta\gamma\omega e^{-\frac{1}{2}\omega^2k_BT/\kappa}\cos(4\pi z_0/\lambda)) D(z_0;dp)/2\pi^2$, while those in Figs.~\ref{fig:Afcsignal}(b, d) correspond to the theoretical cutoff frequency $f_{c,theo}=\kappa D(z_0;d_p)/(2\pi k_BT)$. For parameter identifiability, we fit the data using the following expressions:
\beeq{\label{eq:psd_amplitudefc}
A_{theo}&=\left(\mathcal{A}_{fit}+\frac{\mathcal{B}_{fit}}{2z_{\rm stage}+q_{fit}}\cos\left(\frac{4\pi}{\lambda_{fit}}z_{\rm stage}+\phi_{0,fit}\right)\right)\times\\& \ \ \frac{1}{2\pi^2}\left(1-\frac{9}{16}\frac{d_{fit,1}}{z_{\rm stage}-\delta_{z,1}}+\frac{1}{16}\frac{d_{fit,1}^3}{(z_{\rm stage}-\delta_{z,1})^3}\right)\\
f_{c,theo}&=\mathcal{C}_{fit} \left(1-\frac{9}{16}\frac{d_{fit,2}}{z_{\rm stage}-\delta_{z,2}}+\frac{1}{16}\frac{d_{fit,2}^3}{(z_{\rm stage}-\delta_{z,2})^3}\right),
}
where
\beeq{\mathcal{A}_{fit}&=\beta^2D_0,\\
\mathcal{B}_{fit}&=2\beta\gamma_0\omega e^{-\frac{1}{2}\omega^2k_BT/\kappa}D_0,\\
\text{and} \ \ \mathcal{C}_{fit}&=\frac{1}{2\pi}\frac{D_0}{k_B T}\kappa.
}

The results are summarised in Table~\ref{tab:tablefit}. The distance from the centre of the particle to the cover glass can be estimated with $z_{0,1}=z_{\rm stage}-\delta_{z,1}$ or $z_{0,2}=z_{\rm stage}-\delta_{z,2}$. Note also that the distance between the focal points in the original model ($q$ in Eq.~\eqref{eq:powersignal}) is $q_{fit}+2\delta_{z,1}$  or $q_{fit}+2\delta_{z,2}$; similarly, the phase $\phi_0$ is related to $\phi_{0,fit}+4\pi \delta_{z,1}/\lambda_{fit}$ or $\phi_{0,fit}+4\pi \delta_{z,2}/\lambda_{fit}$. Several fitted parameters, such as the particle diameter and the laser wavelength in the medium, can be compared directly with manufacturer specifications or independent estimates. As shown below, $\mathcal{C}_{fit}$ is directly related with the trap stiffness, and the relationship between $\mathcal{B}$ and $\mathcal{A}$ reflects the contribution of the field scattered forwards to the total signal.   

\begin{table*}[ht!]
\begin{center}
\begin{tabular}{|c|c|c|c|c|c|c||c|c|c|c|}
\hline
   $d_{fit,1}$ & $\mathcal{A}_{fit}$ & $\mathcal{B}_{fit}$ & $\delta_{z,1}$ & $\lambda_{fit}$ & $q_{fit}+$&$\phi_{0,fit}+$&  $d_{fit,2}$ &$\mathcal{C}_{fit}$ & $\delta_{z,2}$& $\kappa_{fit}$\\
     $\left(\mu\mathrm{m}\right)$ & $\left(V^2  s^{-1}\right)$ & $\left(V^2 \mu m s^{-1}\right)$ & $(\mu \mathrm{m})$ & $(\mu \mathrm{m})$ & $2\delta_{z,1}\,(\mu \mathrm{m})$&$\frac{4\pi\delta_{z,1}}{\lambda_{fit}}\,(rad)$ & $(\mu\mathrm{m})$&$\left(s^{-1}\right)$ & $(\mu\mathrm{m})$& $\left(\frac{p\mathrm{N}}{\mu\mathrm{m}}\right)$\\
   \hline
 $0.95$&$0.770$&$0.29$&$-0.15$&$0.819$&$7.1$&$0.42$&$0.99$& $220.8$&$-0.35$&$12.3$\\

 $\pm 0.02$&$\pm 0.001$&$\pm 0.05$&$\pm 0.02$&$\pm 0.001$&$\pm 2.4$& $\pm 0.39$&$\pm 0.06$& $\pm1.3$&$\pm 0.07$&$\pm0.4$ \\
  \hline
 $0.49$&$0.110$&$0.19$&$-0.32$&$0.827$&$4.1$&$0.28 $&$0.49$ & $130.3$&$-0.35$& $3.6$\\
 $\pm 0.06$&$\pm 0.0007$&$\pm 0.02$&$\pm 0.09$&$\pm 0.001$&$\pm 0.9$&$\pm1.5$&$\pm 0.08$&$\pm1.3$ &$\pm 0.13$& $\pm 0.1$\\
 \hline
\end{tabular}
\caption{Fitting parameters obtained from the model in Eq.\eqref{eq:psd_amplitudefc} corresponding to the data shown in Fig.~\ref{fig:Afcsignal}. The first row lists the parameters obtained for the particle with diameter $d_p=0.96\pm0.03\mu\mathrm{m}$, and the second row corresponds to the particle with $d_p=0.50\pm0.02\mu\mathrm{m}$. The laser wavelength in the medium is $\lambda=\lambda_0/n_m =0.8159\,\mu \mathrm{m}$, where $n_m=1.3237$ is the refractive index of water  at $\lambda_0=1.080\,\mu \mathrm{m}$ \cite{thormahlen1985refractive}. The last column shows an estimation of the stiffness using $\mathcal{C}_{fit}$ following Eq.~\eqref{eq:stiff2}.}\label{tab:tablefit}
\end{center}
\end{table*}

\begin{figure}[ht!]
    \centering
\includegraphics[width=0.7\linewidth]{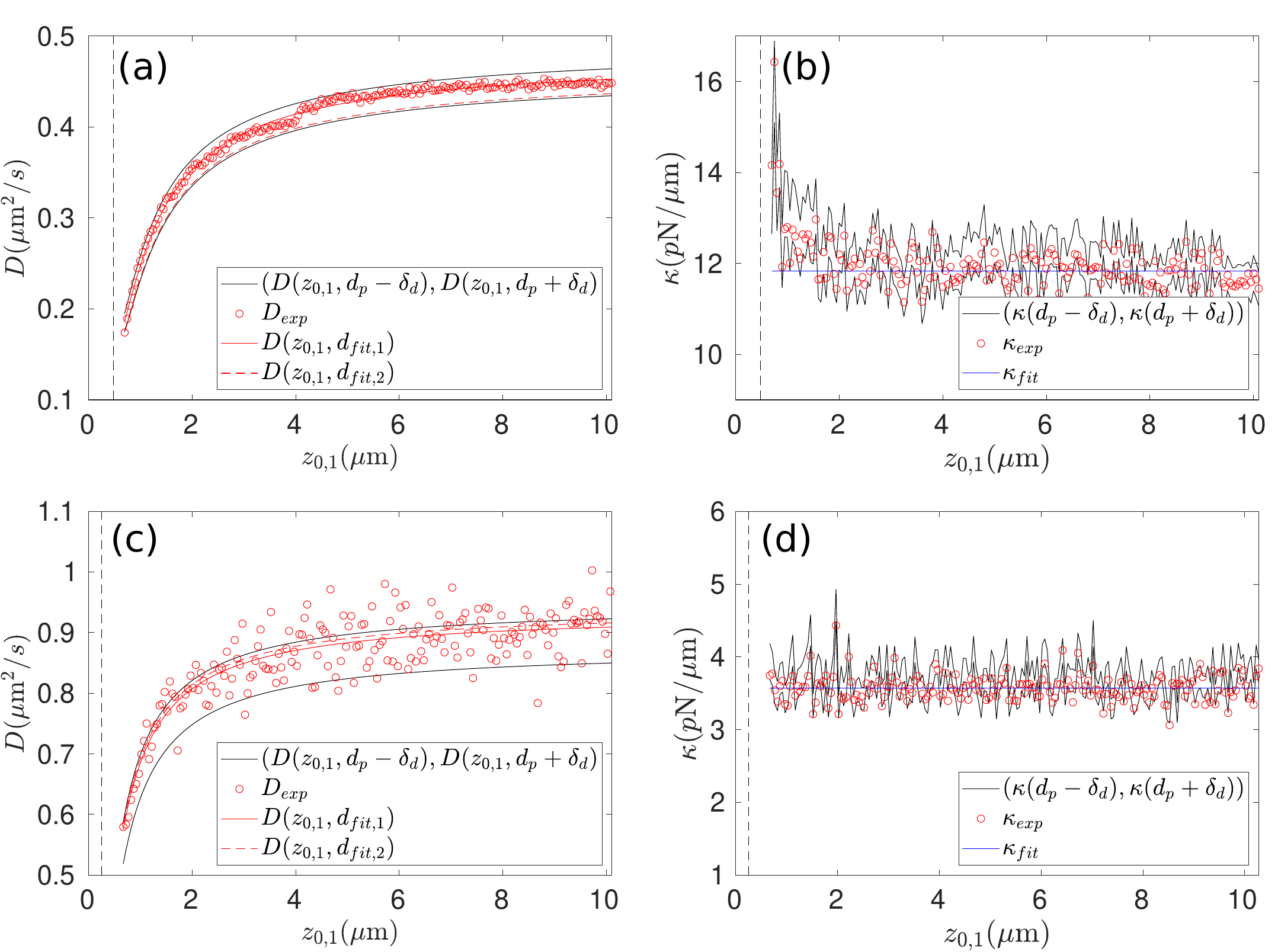}
    \caption{Diffusion coefficient ($D$) and trapped stiffness ($\kappa$) for an OT with particles of diameters $d_p=0.96\,\mu\mathrm{m}$ (a,b) and $0.50\,\mu\mathrm{m}$ (c,d). Based on the experimetnal data  and the model-fitting parameters summarized in Table~\ref{tab:tablefit}, several approaches were used to estimate these quantities (see main text for details). The black solid lines correspond to calculations using the lower and upper limits of the particle diameter specified by the manufacturer, incorporating Faxen's correction and assuming $z_0=z_{0,1}$ (Eqs.~\eqref{eq:faxen} and \eqref{eq:stiff0}). Circles denote experimental estimates according to Eqs.~\eqref{eq:modelfit1} and \eqref{eq:stiff1}. The red dashed and solid lines in (a) and (c) were obtained using Faxen's correction with $z_0=z_{0,1}$ and the particle diameters obtained from the fitting parameters reported in Table~\ref{tab:tablefit}, $d_{exp,1}$ and $d_{exp,2}$. Vertical dashed lines indicate the distance from the cover glass equal to the radius of the particle.}
    \label{fig:diffstiff}
\end{figure}

The particle diffusion coefficient and stiffness of the optical trap are estimated using different approaches. The diffusion coefficient incorporating Fax\'en's correction, given by Eq.~\eqref{eq:faxen}, was evaluated using $d_{fit,1}$ and $d_{fit,2}$ (shown in Table~\ref{tab:tablefit}), i.e. we evaluated $D(z_{0,1},d_{fit,1})$ and $D(z_{0,1};d_{fit,2})$, shown as solid and dashed red lines in Figs.~\ref{fig:diffstiff}(a, c). A different approach uses the full experimental dataset and fitted parameters via
\beeq{\label{eq:modelfit1}
D_{exp}=\frac{2\pi^2A_{exp} D_{0,1} }{\mathcal{A}_{fit}+\frac{\mathcal{B}_{fit}}{2 z_{0,1}+q_{fit}}\cos(4\pi z_{0,1}/\lambda_{fit}-\phi_{0,fit})},
}
where $D_{0,1}=k_B T/(3\pi\nu d_{fit,1})$. The resulting position-dependent values are shown as red circles in Fig.~\ref{fig:diffstiff} (a,c). As reference bounds, we also computed  Fax\'en's diffusion  at the lower and upper limits of the manufacturer-reported diameter, i.e. $D (z_{0,1}, d_p\pm\delta_d$). These comparisons show that the model accurately reproduces the expected position-dependent diffusion.

Alternatively, the stiffness of the trap can be estimated from the cutoff frequency, assuming the diffusion is known:
\beeq{\label{eq:stiff0}
\kappa (d_p\pm \delta_d)=2\pi \frac{k_BT}{ D(z_{0,1},d_p\pm\delta_d)}f_{c,exp}.}
 Another approach uses experimental estimates of both the diffusion and the cutoff frequency:
\beeq{\label{eq:stiff1}
\kappa_{exp}=2\pi \frac{k_BT}{D_{exp}} f_{c,exp}. 
}
The simplest way is to  use the fitted value of $\mathcal{C}_{fit}$ and the bulk diffusion 
\beeq{\label{eq:stiff2}
\kappa_{fit}=2\pi k_BT \mathcal{C}_{fit}/D_{0,1},}
which yields a constant value. The estimates based on these expressions can be seen in 
Figs.~\ref{fig:diffstiff}(b,d) with black lines, red circles, and blue lines, respectively. We observe that $\kappa_{exp}$ is approximately constant and consistent with $\kappa_{fit}$. Estimates based only on the cutoff frequency (Eq.~\eqref{eq:stiff0}) are less accurate, exhibit larger fluctuations, and, for the largest particle, show position-dependent behaviour.

\section{Effect of power on the detection}
\begin{figure}[ht!]
    \centering
\includegraphics[width=0.7\linewidth]{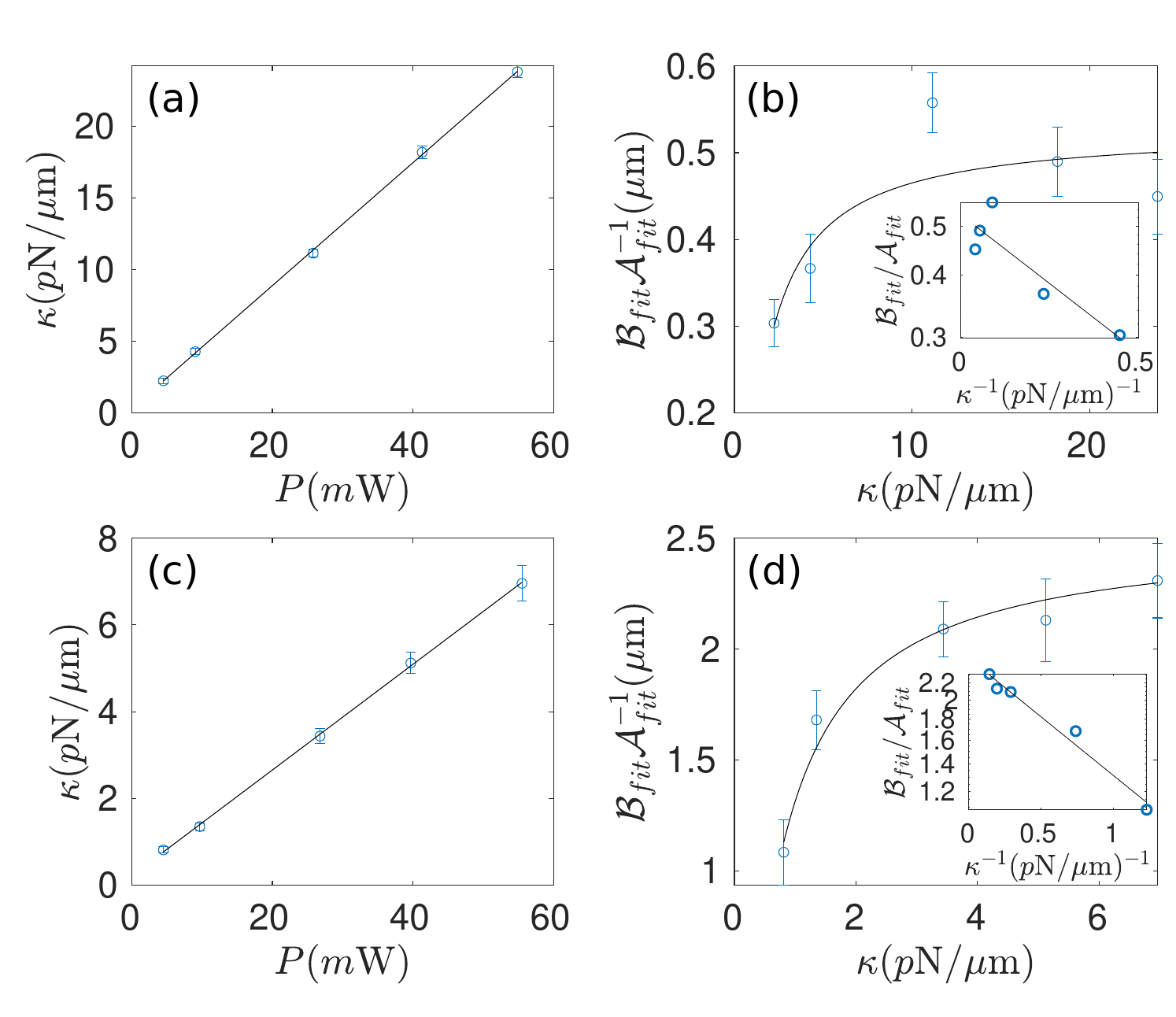}
    \caption{Effect of power and stiffness on the relative amplitude of the oscillations ($\mathcal{B}/\mathcal{A}$) for particles with diameter $d_p=0.96\,\mu\mathrm{m}$ (a,b), and  $d_p=0.5\,\mu\mathrm{m}$ (c,d). (a) and (c) show the linear dependence of the trap stiffness on the optical power (blue dots), estimated using Eq.~\eqref{eq:stiff1}.  The dark solid lines are the linear fits to data points, giving the slopes $\Delta \kappa/\Delta P=0.42\,p\mathrm{N} /\mu\mathrm{m}m\mathrm{W}$ for the $0.96\,\mu\mathrm{m}$ particle and $\Delta \kappa/\Delta P=0.13\,p\mathrm{N}/\mu\mathrm{m}m\mathrm{W}$ for the $0.5\,\mu\mathrm{m}$ particle.  (b) and (d) show the ratio $\mathcal{B}/\mathcal{A}$ as a function of the stiffness determined in (a) and (c). The dark solid lines in these plots are the fits to the exponential function $\mathcal{B}/\mathcal{A}=2\gamma_0 w e^{-w^2k_B T/2\kappa}/\beta$ with free parameters $2\gamma_0 w/\beta$ and  $w^2k_B T/2$. The insets in (b) and (d) highlight the exponential dependence by plotting the same data  as a function of the inverse of the stiffness on a semilogarithmic (y-axis) scale. From these fits, we can estimate the effective Rayleigh range $z_{R,eff}=\lambda/(w\lambda-4\pi)=110\pm50\,n\mathrm{m}$ and $ 400\pm150\,n\mathrm{m}$ for the $d_p=0.96\,\mu\mathrm{m}$ and  $0.5\,\mu\mathrm{m}$ particles, respectively.}
    \label{fig:pow}
\end{figure}

By probing the detection signal at different trapping powers, we observe, as expected, that the estimated trap stiffness varies linearly with power. Figures~\ref{fig:pow}(a, c) show this behaviour for both particle sizes. 
However, the prominence of the standing-wave component depends on the ratio $\mathcal{B}/\mathcal{A}$, which gives the relative weight of the oscillatory term compared with the linear term at a fixed position $z_0$. From the PSD model (Eq.~\eqref{eq:psdmodel}), together with Eq.~\eqref{eq:psd_amplitudefc}, $\mathcal{B}/\mathcal{A}=2\gamma_0 w e^{-w^2k_B T/2\kappa}/\beta$. Since $\gamma_0$ and $\beta$ scale linearly with power, the non-linear power dependence of the signal arises from the exponential factor, which depends on the inverse stiffness. This behaviour is illustrated in Figs.~\ref{fig:pow}(b, d). From these fits, we obtain $w=25\,\mu m^{-1}$ for the big particle and $18\,\mu m^{-1}$ for the small one. With this, and following the discussion in sec.~3, we estimate the corresponding effective trap ranges, $z_{R,eff}=110\,n\mathrm{m}$ and $400\,n\mathrm{m}$. We note that the effective Rayleigh range $z_{R,eff}$ for the smaller particle is comparable to the beam Rayleigh range ($z_R=0.316\,\mu\mathrm{m}$ for an ideal focused beam, as it is described in sec.~2), which is consistent with subwavelength particles behaving as point dipoles. From the same fits in Figs.~\ref{fig:pow}(b, d), we also estimate the ratio of backward- to forward-scattered contributions to the total interference signal as \beeq{\label{eq:ratiopow}
\frac{P_{\rm BS}}{P_{\rm FS}}=\frac{\gamma_0}{2z_0+q}\frac{1}{\beta z_R},}
giving $P_{\rm BS}/P_{\rm FS}=0.10\times 10^{-6}/(2z_0+q)$ and $0.17\times 10^{-6}/(2z_0+q)$ for particle diameters $0.96\,\mu\mathrm{m}$ and $0.50\,\mu\mathrm{m}$, respectively. Thus, the backward-scattered contribution is approximately 1.7 times larger for the small particle than for the large one. Moreover, at separations $2z_0+q$ on the order of a micron, this contribution is not negligible, whereas it becomes negligible several micrometres farther away. Interestingly, according to Eq.~\eqref{eq:ratiopow}, increasing the separation between the focal planes of the objective and the condenser ($q$) would reduce the standing wave effect.

\section{Discussion and conclusions}
In this work, we develop a minimal yet comprehensive model to describe the detection signal in a standard OT configuration. The model explicitly includes the contribution of the backward-scattered light reflected by the bottom cover glass, which adds to the total interference field collected by the condenser. This contribution leads to periodic modulation of the mean, width, and overall shape of the signal's probability distribution as a function of the distance between the particle and the cover glass. These effects were experimentally demonstrated using two silica particles of different sizes, representative of those commonly employed in optical-tweezers experiments.

Despite the inherently nonlinear behaviour of the detected signal, we showed that its power spectral density is accurately described by a Lorentzian profile characterised by a well-defined cutoff frequency and an amplitude that oscillates with the particle-surface distance. This enables reliable estimation of both the diffusion coefficient and trap stiffness through model fitting, while retaining compatibility with standard PSD-based calibration methods. Based on the model and the experimental realisations, we can also accurately infer other relevant system parameters-such as particle size, the wavelength of the laser in the medium, the relative contributions of the backward and forward scattering to the total signal, and the position of the condenser's focal plane-among others.

Beyond its immediate relevance for accurate OT calibration, this model provides a unified framework to quantify surface-induced interference effects in optical detection. It can be extended to explore the influence of refractive-index mismatches or complex geometries, particularly when probe beads are smaller than the wavelength, which is relevant in high-resolution biophysical measurements and near-surface photonic force microscopy. More generally, our results emphasise the importance of accounting for backscattering contributions when interpreting fluctuation-based measurements in optical manipulation systems.

\section{Supplementary information}
\subsection*{Methods and materials}
The sample cell was made with two clean cover glasses ($\#$1.5 Menzel-Gläser and Epredia) and parafilm side bands as spacers. The array was then heated at 50$^o\,C$ to slightly melt the parafilm. The resulting cell was filled with a colloidal suspension of diluted particles (Microparticles GmbH, silica particles with diameters $d_p=0.96\pm 0.03\,\mu\mathrm{m}$ and $0.50\pm0.02\,\mu\mathrm{m}$) in deionised water (Milli-Q) and then sealed with epoxy glue.  
For the optical setup (Fig.~\ref{fig:scheme}), the following brands and models of the materials were used: the laser source
Azurlight Systems ALS-IR-1064-3-SF (3W CW, polarization 100:1. TEM00 M2$<$1.1,  single freq. $<$1MHz, nominally emitting in the range 976-1080$\,n$m); the objectives (OL and CL) Zeiss C-APOCHROMAT 63X (NA=1.2, water immersion, UV-VIS-IR); the piezo stage (PS) PI Physik instrumente E-712.3, the plano convex lenses Thorlabs LA1301-C (L1, with focal length f=250 mm) and Thorlabs LA1805-C (L2, with focal length f=30 mm); the photodiode (PD) Thorlabs DET10N2; and the FPGA card (FPGA) National Instruments PXI-7852R. For the objective and condenser, we used immersion oil with refractive index 1.334 (Zeiss immersol w 2010). 

\subsection*{Dynamics and statistical estimators}
At equilibrium, the trapped particle experiences a harmonic potential and therefore follows the Boltzmann distribution, resulting in a normal distribution with variance $\sigma^2=k_B T/\kappa$.

The overdamped dynamics of the particle along the $z$ axis of the optical trap is described by the Langevin equation
\beeq{\label{eq:langevin}
\dot{z}=-\frac{\kappa_z}{\eta} z+\sqrt{2 D} w(t),
}
where $\kappa_z$, $\eta$ and $D = k_B T/\eta$ denote, respectively, the stiffness of the trap along $z$, the drag coefficient and the diffusion coefficient. The stochastic term $w(t)$ represents  delta-correlated (white) thermal noise, satisfying $\langle w(t) w(t')\rangle=\delta(t-t')$. 

At equilibrium, the particle position follows a Gaussian distribution,
\beeq{\label{eq:dist1}
\rho(z|\sigma^2,\mu)=\frac{1}{\sqrt{2\pi\sigma^2}}\exp\left(-\frac{(z-\mu)^2}{2\sigma^2}\right)\,,
}
with mean $\mu=0$ and variance $\sigma^2=k_\mathrm{B}T/\kappa_z$, where $k_B$ is the Boltzmann constant and $T$ the  absolute temperature. 

The autocorrelation function (ACF) and the power spectral density (PSD) of the dynamics are expressed as
\beeq{\label{eq:acfpsd}
\mathrm{ACF}(t)&=\sigma^2e^{-t/\tau}\,,\quad  \mathrm{PSD}(f)=\mathcal{F}\{\mathrm{ACF}(t)\}=\frac{1}{2\pi^2}\frac{D}{f_c^2+f^2},
}
where $\tau=\eta/\kappa_z$ is the characteristic time of the trap.

Experimentally, both the ACT and PSD can be directly computed from the time series of particle position, enabling robust estimation of the stiffness of the trap and the diffusion coefficient and, consequently, precise calibration of the optical-tweezers system \cite{Gieseler:21}. 

\subsection*{Out of equilibrium case}
We now consider a Brownian particle trapped by optical tweezers in a non-equilibrium state. In this regime, the  probability of observing a  trajectory  $\{z_n\}_{n=1}^T$ at discrete times, with  $z_n=z(t_n)$, is Gaussian and can be written as:
\beeq{
P(\{z_n\}_{n=1}^T)=\frac{1}{\sqrt{(2\pi)^T\det C}}\exp\left[-\frac{1}{2}\sum_{n,\ell=1}^T(z_n-\mu_n)C^{-1}_{n\ell}(z_\ell-\mu_\ell)\right]\,
}
where $\mu_n=\bracket{z_n}$ and $C_{n\ell}=\bracket{(z_n-\mu_n)(z_\ell-\mu_\ell)}$ are the mean and covariance of the process:
\beeq{
\mu_n=z_0e^{-t_n/\tau}\,,\quad\quad C_{nm} = D \tau\left[ e^{-|t_n-t_m|/\tau}-e^{-(t_n+t_m)/\tau}\right]\,,
}
where $\tau=\frac{\gamma}{\kappa}$. notice that $\mu_n$ tends to cero in the long-term, or stationary, regime.

The detected signal is related to the position of the particle through the nonlinear mapping
\beeq{
s_n=\alpha+\beta z_n+\gamma \sin(a z_n+b),
}
which corresponds to our minimal model generalising the linear response discussed in the main text.

To characterise the statistical properties of the measured signal, we derive the two-point autocorrelation function of $S$. This requires computing the first and second moments,
\beeq{
\bracket{s_n}&=\alpha+\beta \bracket{z_n}+\gamma \bracket{\sin(a z_n+b)}\,,\\
\bracket{s_n s_\ell}&=\bracket{\left(\alpha+\beta z_n+\gamma \sin(a z_n+b)\right)\left(\alpha+\beta z_\ell+\gamma \sin(a z_\ell+b)\right)}\,.
}

Because the underlying distribution is Gaussian, the averages follow standard manipulations based on the moment-generating (or characteristic) function of multivariate normal distributions. Specifically, the calculation requires expanding the product of the non-linear mapping appearing in the second moment and evaluating moments of the form $\bracket{z_n z_\ell}$, $\bracket{z_n \sin(a z_\ell + b)}$, and $\bracket{\sin(a z_n + b)\sin(a z_\ell + b)}$. These terms can be obtained analytically using well-known Gaussian integration identities. Because the intermediate steps are straightforward and follow standard results, we only report here the final expressions for the first two moments. These are:

\beeq{
\bracket{s_n}&=\alpha+\beta \mu_n+\gamma e^{-\frac{a^2 C_{nn}}{2}}\sin(a \mu_n+b)\,,\\
\bracket{s_ns_\ell}&=\alpha^2+\alpha\beta\mu_n+\alpha\beta\mu_\ell+\beta^2[C_{n\ell}+\mu_\ell\mu_n]+\\
&\gamma^2\Bigg[\frac{1}{2}e^{-\frac{a^2 C_{nn}}{2}-\frac{a^2 C_{\ell\ell}}{2}+a^2C_{n\ell}}-\frac{1}{2}e^{-\frac{a^2 C_{nn}}{2}-\frac{a^2 C_{\ell\ell}}{2}-a^2C_{n\ell}}\cos(2b)\Bigg]\\
&+\alpha\gamma e^{-\frac{a^2 C_{nn}}{2}}\sin(a \mu_n+b)+\alpha\gamma e^{-\frac{a^2 C_{\ell\ell}}{2}}\sin(a \mu_\ell+b)\\
&+\beta\gamma \Bigg[\mu_\ell e^{-\frac{a^2C_{nn}}{2}}\sin(a\mu_n+b)+aC_{n\ell}e^{-\frac{a^2C_{nn}}{2}}\cos(a\mu_n+b)\Bigg]+\\
&\beta\gamma \Bigg[\mu_n e^{-\frac{a^2C_{\ell\ell}}{2}}\sin(a\mu_\ell+b)+aC_{n\ell}e^{-\frac{a^2C_{\ell\ell}}{2}}\cos(a\mu_\ell+b)\Bigg]
}
Combining these results, we have the following expression for the two-point connected auto-correlation function:
\beeq{\mathrm{ACF}_s=
\bracket{s_ns_\ell}-\bracket{s_n}\bracket{s_\ell}&=\beta^2C_{n\ell}+\\
&\gamma^2\Bigg[\frac{1}{2}e^{-\frac{a^2 C_{nn}}{2}-\frac{a^2 C_{\ell\ell}}{2}+a^2C_{n\ell}}-\frac{1}{2}e^{-\frac{a^2 C_{nn}}{2}-\frac{a^2 C_{\ell\ell}}{2}-a^2C_{n\ell}}\cos(2b)\Bigg]\\
&+\beta\gamma aC_{n\ell}e^{-\frac{a^2C_{nn}}{2}}\cos(a\mu_n+b)+\beta\gamma aC_{n\ell}e^{-\frac{a^2C_{\ell\ell}}{2}}\cos(a\mu_\ell+b)\\
&-\gamma^2 e^{-\frac{a^2 C_{nn}}{2}-\frac{a^2 C_{\ell\ell}}{2}}\sin(a \mu_n+b)\sin(a \mu_\ell+b)\,.
}

\section*{Data availability}
The data that support the findings of this study are available upon reasonable request from the authors.

\section*{Competing interests}
  The authors declare that they have no competing interests.
 
\section*{Acknowledgements}
 
\section*{Funding}
A. V. A acknowledges financial support from UNAM, DGAPA-PAPIIT (project IN104924) and PASPA-DGAPA. S.L. was supported by the UChicago-CNRS Joint PhD Research Program, which is a part of the IRC Discovery. S.D. and A.L.N were supported by the ANR PHYBION (ANR-21-CO15-0004) project grant, as well as by the Impulscience project PHYBION of the Bettencourt Schueller Foundation. 
F.P. was supported by the ANR BaElmec ANR-23-CE30-0010 project grant.
Isaac Pérez Castillo acknowledges financial support from SECIHTI under the research grant CBF-2025-I-3911.
The CBS is a member of the France-BioImaging (FBI) and the French Infrastructure for Integrated Structural Biology (FRISBI), two national infrastructures supported by the French National Research Agency (ANR-10-INBS-04-01 and ANR-10-INBS-05, respectively).


\bigskip
\bibliographystyle{unsrt}
\bibliography{docbiblio, references-zotero}

\end{document}